# Inelastic X-ray scattering studies of phonon dispersions in superconductors at high pressures

**S.M. Souliou[1], A. Bosak[2], G. Garbarino[2] and M. Le Tacon[1]**

[1] Institute for Quantum Materials and Technologies, Karlsruhe Institute of Technology, Karlsruhe, Deutschland
[2] European Synchrotron Radiation Facility, BP 220, F-38043 Grenoble Cedex, France

E-mail: michaela.souliou@kit.edu, matthieu.letacon@kit.edu

## Abstract

Electron-phonon interaction is of central importance for the electrical and heat transport properties of metals, and is directly responsible for charge-density-waves or (conventional) superconducting instabilities. The direct observation of phonon dispersion anomalies across electronic phase transitions can provide insightful information regarding the mechanisms underlying their formation. Here, we review the current status of phonon dispersion studies in superconductors under hydrostatic and uniaxial pressure. Advances in the instrumentation of high resolution inelastic X-ray scattering beamlines and pressure generating devices allow these measurements to be performed routinely at synchrotron beamlines worldwide.

Keywords: high pressure, phonon dispersion, superconductivity

## 1. Introduction

Pressure is commonly used to tune or induce superconductivity. Whereas only one third of the elements of the periodic table are superconducting at ambient pressure, more than half of them exhibit superconductivity under high pressure [1]. In correlated electron systems, pressure is used to tune the balance between the Coulomb repulsion and the electronic bandwidth, which in turn drives the system through various electronic phases that often encompass unconventional or high temperature superconductivity. More recently the world highest superconducting transition temperatures have been obtained under extreme pressures [2-4].

Each time, a key question to answer is that of the nature of the interplay between the different electronic phases of the pressurized material and of the microscopic mechanism yielding the formation of the superconducting condensate. In principle, a general strategy to address this issue consists in studying in parallel the crystal and electronic structures of the material of interest alongside their collective excitations spectra (e.g. phonons, magnons, …) as function of energy and momentum and across the various phase transitions. This then allows to determine which of these excitations display a sizeable coupling to the conduction electrons, and might thereby be directly involved in the formation of one or the other phase.

However, such strategy can rapidly become tedious to put in practice for complex materials, particularly when the main tuning parameter to reach the desired electronic phase is pressure. The experiments that are typically performed on superconductors at high pressures are electrical transport or thermodynamic (e.g. susceptibility) measurements as well as diffraction studies of the crystal structure. The highest pressures are achieved using diamond anvil cells (DAC) which are particularly suitable for x-ray diffraction at synchrotron radiation facilities [5]. The DAC environment is also compatible with nuclear magnetic resonance and particularly suitable for light scattering or infrared spectroscopy studies of superconducting materials[6]. These are however limited to zone-center ($Q = 0$) excitations by the small momentum of visible photons.

In this article, we briefly review some recent experimental work that took advantage of the progress in the inelastic x-ray scattering (IXS) technique to investigate the effects of hydrostatic or uniaxial pressure on the phonons of various families of superconductors.

The electron-phonon interaction (EPI) plays a central role in the electrical and thermal properties of solids [7], and is primarily involved in a certain number of collective phenomena, starting with superconductivity in conventional superconductors, in which it drives the Cooper pairing, or with the formation of charge-density-waves (CDW) in low-dimensional metallic systems. In many compounds, encompassing metallic dichalcogenides or high temperature superconducting cuprates, both superconductivity and CDW are commonly encountered and naturally involve EPI, but its actual role remains relatively controversial. Fresh insights on this issue were offered by the response of the lattice dynamics upon the pressure-induced suppression or enhancement of the CDW and superconducting orders.

This article is organized as follows. In the next section, we detail the experimental challenges and limitations related to high pressure IXS experiments: compatible high pressure environments (hydrostatic and uniaxial pressures) and sample preparation techniques. Then we review selected results of high pressure IXS investigations on elemental superconductors, transition metal dichalcogenides and finally high temperature superconducting cuprates.

## 2. IXS under high pressure: Experimental challenges

A recent review of the IXS technique and of the currently available experimental stations can be found in reference[8]. Here we will focus on the aspects which are relevant to the high pressure sample environment used for IXS experiments on single crystals.

### 2.1 Pressure generating devices

Although some early studies were done using large volume cells of the piston-cylinder type [9], most IXS experiments under hydrostatic pressure (HP) conditions – and almost exclusively the ones on superconductors – are carried out using the DAC apparatus [10]. The IXS experiments are conventionally performed in transmission geometry through the diamond anvils. Radial scattering geometries using the gasket as an x-ray window[11] are avoided in the energy range of interest due to the overlapping phonon scattering signal from the gasket material (see for instance reference [12] for the beryllium gaskets commonly used in panoramic DACs). In the working geometry, the beam path through the diamonds is ~50 times longer than the path through the sample. Phonon and Bragg scattering signal from the diamonds inevitably appears and often obstructs the measurements. However, it can be circumvented by rotating the DAC around the compression axis and working on structurally equivalent positions in the reciprocal space of the investigated sample [13]. Parasitic signal can also arise from the pressure transmitting medium (PTM), although this is more rarely encountered. The use of PTM is crucial to ensure the best possible hydrostatic conditions and the preservation of the sample's crystalline quality throughout the measurements. The choice of the PTM in IXS experiments is governed by the standard criteria for chemical inertness to the sample and high hydrostaticity at the working temperature range [14]. In most of the reported HP experiments on superconductors, condensed gases were employed as PTM. Amongst them, helium is considered to be the best choice, particularly for measurements at low temperatures.

Recently, novel piezoelectric–based straining devices have been developed for the transport or thermodynamic studies of uniaxially pressurized samples [15-17]. To account for the specific needs of photon scattering experiments (*e.g.* transmission geometry), dedicated strain cells were developed and used in IXS experiments [18, 19].
They combine a compact design and an extended, unobstructed angular access of the incident (scattered) beam to (from) the strained sample. Their basic working principle relies on controlling the displacement of two moving blocks by varying the voltage across the piezoelectric actuators. The sample under investigation is shaped in the form of a thin bar and firmly mounted with an epoxy across the two moving blocks. Through a suitable variation of the bias voltage, the sample can be compressively or tensily stressed in a continuous and well-controlled manner and with high strain homogeneity at its central part. The applied strain can be obtained by measuring the strained lattice parameters by x-ray diffraction, but also by using a capacitive strain gauge positioned parallel and underneath the mounted sample. The capacitor is kept in a feedback loop, therefore serving as a guarantee of constant strain on the sample during the typically long IXS measurements.

### 2.2 Cryogenic temperatures

In the study of superconductors high pressures are often necessarily coupled to cryogenic temperatures. While the combination of high pressure and low temperature conditions adds up to the technical complexity, these experiments open the route for an extended (P, T) mapping of the phase diagrams and often enable reaching otherwise inaccessible phases of the studied samples. Dedicated cryostats are typically tailor-made for specific types of DACs. These are developed to meet the requirements for compactness, large window openings, geometrical flexibility, precise motorized translation and rotation, as well as remote pressure control at low temperatures. Optical access is very often also envisaged, allowing the use of the ruby fluorescence lines for the pressure calibration inside the DAC [20].

In addition to the above, in the case of piezoelectric-based uniaxial straining devices another parameter to consider at cryogenic temperatures is the differential thermal contraction between the actuators and the strained sample. The straining cell used for IXS experiments in reference [19] addressed this issue combining a set of compression and tension piezoelectric stacks compensating each other upon cooling.

### 2.3 Samples for high pressure experiments

Sample preparation and handling is a crucial and challenging task for IXS measurements under HP. This is not only because of the small sample size required to fit the DAC sample chamber (typical sample diameter <100 μm and thickness ~30 μm), but also due to the limited accessible part of the reciprocal space imposed by the DAC design (angular opening of the diamonds ~60°). For this reason, it is essential to prepare and load the sample under investigation in a suitable pre-selected orientation which guarantees that the reciprocal space parts of interest rest accessible. It is worth noting that the sample is not glued or mechanically fixed inside the DAC chamber, remaining therefore free to rotate, for instance across the solidification of the surrounding PTM.

Also in the case of uniaxial straining devices, the needle-like geometry requires a multi-step sample preparation procedure. This includes the wire-saw cutting of the studied crystal in bars, followed by a fine mechanical polishing on all four sides in order to achieve a uniform bar cross-section and to avoid surface cracks. For experiments in transmission geometry, the final bar thickness is selected based on the absorption length of the studied sample in the used incident energy.

Gentle cutting processes, such as femtosecond-laser cutting and focused-ion-beam micromachining, are particularly well-suited for sample preparation both for HP and for uniaxial experiments. Such techniques allow reaching the targeted dimensions in any given orientation, while maintaining the sample quality required for single crystal IXS experiments.

## 3. Selected studies

### 3.1 Elemental Superconductors

Uranium crystallizes in an orthorhombic structure uniquely encountered for an element under ambient pressure conditions (α-U phase, space group *Cmcm*). CDW ordering appears below 43 K and induces a structural modulation (α₁-U) which involves approximately a doubling of the unit cell along the a-axis with respect to the unmodulated α-U structure. The CDW order in uranium was associated to nesting features of its Fermi surface (nesting vector $Q_N = (1/2,0,0)$ in reciprocal lattice units (RLU)). Under pressure application, the CDW disappears, whereas at the same time the superconducting critical temperature $T_c$ is increased. $T_c$ is maximized (~2 K) when the CDW order is completely supressed at ~1.5 GPa.

Raymond et al. investigated the role of EPI in mediating the competition between superconductivity and CDW ordering in uranium through a combination of HP IXS experiments and *ab initio* calculations [21]. The authors have measured the softening of the $\Sigma_4$ phonon branch around $q_N$ at room temperature and observed the gradual pressure-induced suppression of the dispersion dip until its complete disappearance at 20 GPa (see Fig.1). Interestingly, calculations of the electronic structure indicated that the nesting features are rather insensitive to the application of pressure and remain unaltered up to 20 GPa. Nevertheless, calculations of the electron–phonon coupling in the α-U phase revealed a strong suppression under HP. Moreover, the authors showed that - unlike the case of the α-U structure - in the modulated α₁-U structure (up to 1.5 GPa) the electron-phonon coupling increases under pressure, and related this increase to the rise of the superconducting $T_c$. All together, these results highlight that the momentum and pressure dependent EPI plays a dominant role in the observed disappearance of the soft mode and in shaping the overall interplay of superconductivity and CDW order in the phase diagram of uranium.

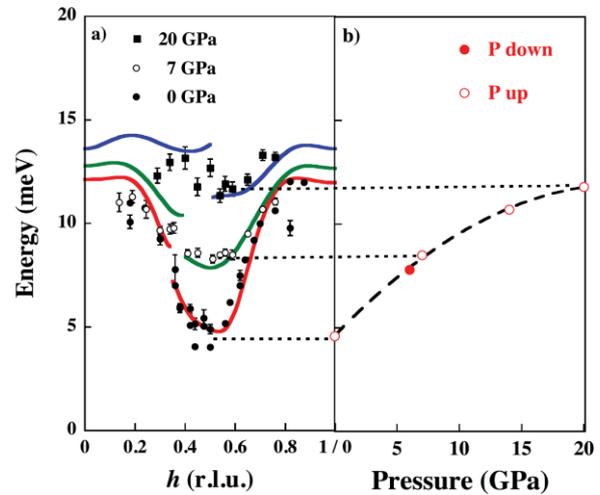

Figure 1 *(a) Experimental (data points) and theoretical (solid lines) dispersion of the $\Sigma_4$ phonon branch along the [100] direction at ambient pressure, 7 GPa and 20 GPa. (b) Soft-mode energy as a function of pressure (lines are guides to the eye). Reprinted figure with permission from [21] Copyright (2011) by the American Physical Society.*

Following the HP IXS investigation on uranium, Loa and co-authors studied the lattice dynamics in the high pressure α′ phase of cerium [22]. The α′ phase is isostructural to the α-U phase of uranium and it is stabilized in elemental cerium above 5-6 GPa. Pronounced anomalies were observed in the phonon dispersions of pressurized cerium, strongly resembling the ones reported for the α-U phase. More specifically, a clear dip of the $\Sigma_1$ and $\Sigma_4$ phonon branches was observed around $Q=(1/2,0,0)$ (Fig. 2). The similarities between the phonon anomalies in the two systems, which share the same crystal structure but otherwise have very different electronic structures, becomes more intriguing when taking into account that up to date there has been no experimental observation of CDW ordering in cerium. Calculations of the Fermi surface have not revealed particularly strong nesting features at the wavevector where the phonon anomalies are seen. Nevertheless, strong electron-phonon coupling was calculated around $Q=(1/2,0,0)$ due to large electron-phonon matrix elements and was suggested to be the origin of the pronounced dips of the $\Sigma_1$ and $\Sigma_4$ phonon dispersions. The strong electron-phonon coupling calculated for this phase in comparison to alternative structures stabilized in elemental cerium under pressure, led the authors to the conclusion that the superconducting phase of cerium is the $\alpha'$ phase.

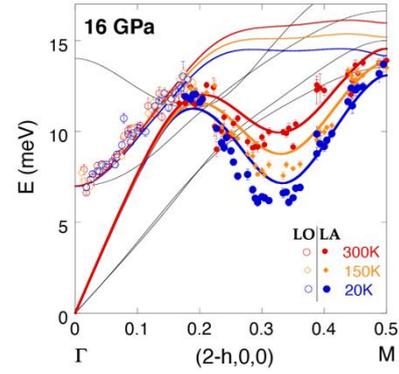

The strong temperature dependence of the soft phonon well beyond the stability range of the CDW indicates that anharmonic effects play an important role and should be considered as a contributing mechanism in shaping the phase diagram of this system. Indeed, harmonic ab-initio calculations fail to describe the ground state of 2$H$-NbSe$_2$ as a function of pressure, as they predict a CDW instability up to 14 GPa. The authors have carried out a series of calculations which explicitly include the effects of anharmonicity (stochastic self-consistent harmonic approximation). The results of these calculations, which are in very good agreement with the experimentally determined pressure and temperature dependence of the phonon dispersions, succeed in correctly predicting the stability range of the CDW.

In the related system 1$T$-TiSe$_2$ CDW order develops at ambient pressure below $T_{CDW}$ = 200 K. IXS experiments and *ab-initio* calculations showed that the CDW transition and the accompanying lattice modulation are related to the softening of a transversely polarized phonon mode at the L point of the high temperature trigonal Brillouin zone [25]. Nevertheless, several experimental observations and theoretical investigations are consistent with an excitonic insulator picture, in which exciton condensation plays a major role in driving the CDW modulation in 1$T$-TiSe$_2$. It has been argued that as excitons necessarily couple to the underlying phonons, the resulting hybrid phonon-exciton modes are the ones mediating the CDW formation [26]. Superconductivity appears in this system either under pressure or upon Cu intercalation. Unlike the case of Cu intercalation, the superconducting dome under pressure is surrounded by the CDW ordered phase rather than being centered on the critical pressure at which CDW order disappears. The interplay

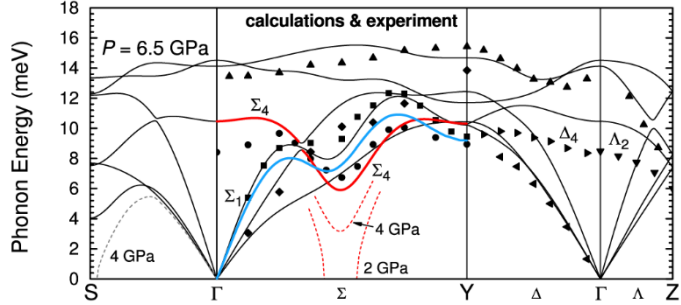

### 3.2 Transition metal chalcogenides

Many members of the transition metal dichalcogenide family exhibit CDW phases with highly diverse characteristics which compete with superconductivity. The phononic contribution to the stabilization of these electronic phases, as well as to the tuning of their interplay by external parameters is intensively investigated. One of the most studied examples is 2$H$-NbSe$_2$, in which incommensurate CDW ordering sets in at $T_{CDW}$ = 33.5 K and coexists with superconductivity below $T_c$ = 7.2 K. The second order transition to the CDW state is related to the softening of a longitudinal acoustic phonon mode. It has been highlighted that the extended $q$-width of the observed phonon dip around the CDW ordering vector $q_{CDW}$ is not in line with the sharply localized dips expected from Fermi surface nesting. Strongly q-dependent electron-phonon coupling has been proposed as the driving force behind the CDW formation and the one determining the ordering wavevector [23].

Leroux et al. have investigated the pressure dependence of the soft phonon in 2$H$-NbSe$_2$ [24]. Earlier HP experiments have showed the pressure-induced suppression of the CDW order, which disappears above ~4.5 GPa, while superconductivity remains almost unaffected. In their HP experiment, the authors observed that the strong temperature dependence of the soft phonon survives up to at least 16 GPa, which is much higher than the critical pressure of ~4.5 GPa above which the CDW is fully suppressed (Fig. 3). At 16 GPa, the amplitude of the temperature dependent phonon softening is certainly reduced, but remains significant as the phonon energy at $q_{CDW}$ decreases by almost a factor of 2 (see Fig. 3).

between the two phases implies that the formation mechanism of the CDW directly relates to the superconducting pairing mechanism. The different effect of chemical intercalation and physical pressure, raised questions on the individual contribution of phonons, excitons and hybrid modes when tuning the interplay of CDW and superconductivity by external parameters.

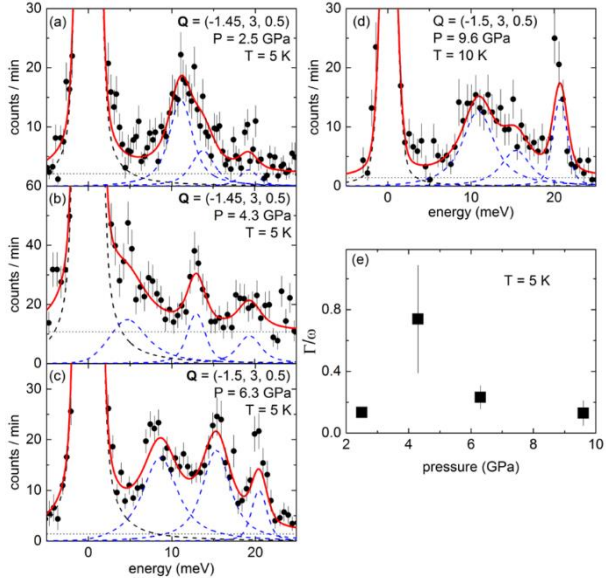

*Figure 4 (a)-(d) IXS spectra measured at 5 K under a pressure of (a) 2.5 GPa, (b) 4.3 GPa, (c) 6.3 GPa and (d) 9.6 GPa. (d) Pressure dependence of the ratio of the soft mode's linewidth Γ to its energy ω. Reprinted figure with permission from [27] Copyright (2016) by the American Physical Society.*

Maschek et al. addressed this issue through a comparative IXS study of $Cu_xTiSe_2$ and pressurized $1T$-$TiSe_2$ combined with density function perturbation theory (DFPT) calculations [27]. According to their DFPT calculations, the soft mode related to the CDW carries the largest part of the total electron-phonon coupling and should be the most relevant for phonon-mediated superconductivity. The authors investigated experimentally the contribution of this phonon to the superconducting $T_c$, by following the ratio of the soft mode's linewidth to its energy, $\Gamma/\omega$. In the case of $Cu_xTiSe_2$ the experimental results show an increase of the ratio at low temperatures upon approaching the soft L point in the superconducting doped compounds. This is in line with what is expected from DFPT calculations when the electron-phonon coupling of the soft phonon mode is the driving force for the superconducting phase near the critical point of the CDW suppression. Interestingly, in the case of pressurized $1T$-$TiSe_2$ the IXS data reveal no dramatic change of the $\Gamma/\omega$ ratio when the superconducting phase is approached under HP (Fig. 4). The authors conclude that –unlike the case of Cu-intercalation– the pressure-induced emergence of superconductivity cannot be solely associated to a phononic soft mode and propose a hybrid phonon-exciton model which can explain the experimental observations. More specifically, Cu intercalation induces a shift of the chemical potential of $1T$-$TiSe_2$, deteriorating the exciton formation conditions and subsequently decreasing the hybrid's mode exciton content. Contrarily, HP application has a relatively small effect on the chemical potential. Considering the crossing between the dispersions of the interacting phonon and exciton modes, the authors show that the low energy hybrid mode reaches zero energy at a higher pressure compared to the bare non-interacting phonon. Therefore, the phonon-exciton hybridization provides an explanation for the shift of the superconducting dome with respect to the critical pressure at which the CDW disappears. Moreover, the authors show within the standard BCS (Bardeen-Cooper-Schrieffer) framework that also the superconducting pairing interaction depends strongly on the degree of phonon-exciton hybridization, which in turn is pressure-dependent as it relies upon the energy difference between the bare exciton and phonon modes.

### 3.3 High-$T_c$ Cuprates

The existence of charge modulations, in the form of stripes, checkerboard or CDWs and their interplay with high temperature superconductivity in the cuprates has been one of the most debated issues in the literature for more than three decades [28, 29]. Anomalies in the dispersion of high energy optical phonons with a displacement pattern similar to that of the expected charge modulations have been discovered for several families of cuprates ($YBa_2Cu_3O_{6+x}$ [30], $La_2(Sr/Ba)CuO_4$ [31], $Bi_2Sr_{2-x}La_xCu_2O_6$ [32]) and interpreted as fingerprints of fluctuating charge stripes [33].

With the more recent discovery of ubiquitous short-range two-dimensional CDW in all the cuprates families [34-39], new investigations of their lattice dynamics have been undertaken and revealed strong temperature dependent anomalies for the lowest energy phonon branches at the incommensurate CDW ordering wave vector $Q_{CDW}$ [40-42]. In underdoped $YBa_2Cu_3O_{6+x}$ (YBCO), where the CDW is best developed with a in-plane correlation lengths up to ~80 Å, the apparition of the CDW at $T_{CDW} > T_c$ is associated with three distinct fingerprints i) a sharp increase of the quasi-elastic line intensity at $Q_{CDW}$ down to $T_c$, which decreases upon further cooling in the superconducting state, ii) a broadening of the lowest energy phonon branches for temperatures down to $T_c$ and iii) a steep sharpening and softening at $Q_{CDW}$ below $T_c$ (also referred to as a superconductivity-induced giant Kohn anomaly). This behavior contrasts a lot with that, more commonly encountered, of soft-mode driven CDWs in low dimensional metals [23,43,44], and indicate an unusual interplay between superconductivity and CDW in these materials.

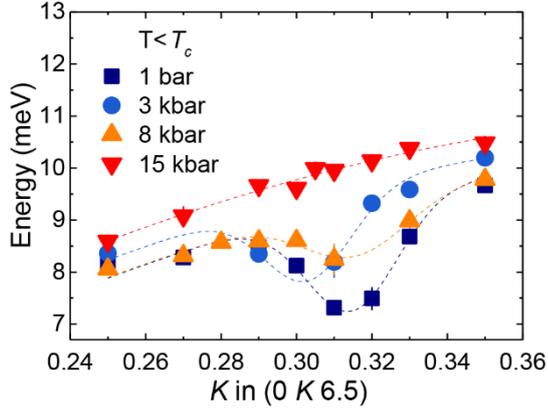

*Figure 5 Pressure and momentum dependence of the acoustic phonon energy below $T_c$ across $Q_{CDW} = (0, 0.315, 6.5)$ in underdoped $YBa_2Cu_3O_{6.67}$. The dashed lines are guides to the eyes. Reprinted figure with permission from [48] Copyright (2018) by the American Physical Society.*

In the underdoped region of YBCO where the CDW is the strongest, hydrostatic pressure results in a steep increase of $T_c$ that almost doubles by 15 GPa [45]. Whereas transport [46] and nuclear magnetic resonance [47] experiments indicate a no or a small suppression of the CDW under pressure, IXS measurement revealed that all the aforementioned low-energy phonon anomalies associated with the CDW at ambient pressure are suppressed by 1 GPa of hydrostatic pressure (Fig. 5) [48]. This suggests a very rapid suppression of the CDW as the superconducting phase is enhanced. The origin of the discrepancy between these experiments, carried out in different types of pressure cells with different PTM remains unclear.

Another approach to study the interplay between superconductivity and CDW consists in suppressing the former, using *e.g.* a magnetic field. To date however, the implementation on IXS beamlines of magnets with sufficient strength to obliterate superconductivity in the cuprates (typically >10T) have not been realized for technical reasons. As an alternative route, it has been established that the dependence of $T_c$ to uniaxial pressure is highly anisotropic. In YBCO in particular, a-axis compression rather than hydrostatic pressure can very effectively suppress superconductivity. Uniaxial compression of the a-axis of underdoped YBCO amounting to ~1 GPa was shown to decrease $T_c$ from ~65 to ~45K [19], to strengthen the 2D CDW and finally to induce a 3D long-range-ordered CDW (Fig. 6), that had been only reached so far using magnetic fields larger than ~15T [49]. This study further revealed that the transition to the 3D CDW is associated with a pronounced, possibly complete, softening of an optical phonon. This illustrates the potential of coupling uniaxial-pressure and spectroscopy to control competing orders and gain insights regarding their formation in the cuprates, and more generally within multi-phase quantum materials.

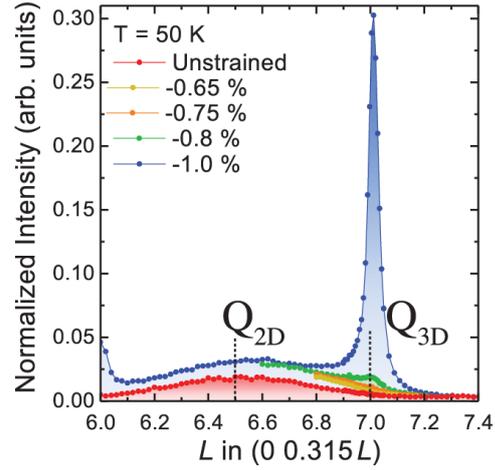

*Figure 6 Strain and momentum dependence of the quasi-elastic line intensity from IXS measurements in underdoped $YBa_2Cu_3O_{6.67}$. at 50 K. From [19]. Reprinted with permission from AAAS.*

## 4. Summary and Outlook

The examples discussed above are representative of the valuable input provided by high pressure phonon dispersion studies in clarifying the role of the EPI in various families of superconductors. These experiments, often combined with *ab initio* calculations, allow to advance our understanding of the EPI contribution in the formation of the superconducting phase and its interplay with other types of electronic ordering, such as the commonly observed CDW order.

Despite the non-negligible technical challenges, in particular when pressure is combined with low temperature conditions, recent technical progress in IXS beamlines and in high pressure techniques have rendered these experiments regularly feasible. Expanding the accessible pressure and temperature conditions is foreseeable in the near future and will allow investigating more superconducting compounds, such as superconducting hydrides and heavy fermion systems, and addressing questions which were so far beyond experimental reach.

## Acknowledgements

We acknowledge fruitful discussions and collaborations with Mark Barber, Alfred Baron, Elizabeth Blackburn, Lucio Braicovich, Nick Brookes, Matteo Calandra, Laurent Cario, Johan Chang, Matteo d'Astuto, Andrea Damascelli, Mark Dean, Tom Devereaux, Ion Errea, Alex Frano, Giacomo Ghiringelli, Hlynur Gretarsson, Martin Greven, Stephen Hayden, Rolf Heid, Clifford Hicks, Jeroen Jacobs, Bernhard Keimer, Hun-ho Kim, Steve Kivelson, Michael Krisch, Wei-Sheng Lee, Maxime Leroux, Yuan Li, Toshinao Loew, Andrew Mackenzie, Marie-Aude Méasson, Matteo Minola, Matteo Mitrano, Marco Moretti, Eduardo da Silva Neto, Juan

Porras, Pierre Rodière, George Sawatzky, Enrico Schierle, Suchitra Sebastian, Wojciech Tabis, Frank Weber, and Eugen Weschke.

MLT acknowledges the funding by the Deutsche Forschungsgemeinschaft (DFG; German Research Foundation) – Project-ID 422213477 - TRR 288.